\documentclass[twocolumn,showpacs,prl,letter]{revtex4}%
\usepackage{amsmath,amssymb,amsthm}
\usepackage{epsfig}
\usepackage{epstopdf}
\usepackage{graphicx}%
\usepackage{amsmath}%
\usepackage{amsfonts}%
\usepackage{amssymb}

\begin{document}

\title{Solitonic Phase Transitions of Galilean Black Holes}
\author{R.B. Mann\footnote{email: rbmann@sciborg.uwaterloo.ca; on leave from
Department of Physics and Astronomy, University of Waterloo,
Waterloo, Ontario, N2L 3G1, Canada} } 
\affiliation{Kavli Institute for Theoretical Physics,
University of California, Santa Barbara, CA 93106, USA}

\begin{abstract}
Planar black holes with Galilean asymptotics have been proposed as the holographic dual of a non-relativistic conformal field theory at finite temperature. I show that these objects can undergo a phase
transition at low temperature to a new type of soliton with the same asymptotic structure. The strength of the chemical potential is induced by the
scale of the soliton. 
  \end{abstract}

\maketitle

Holography is broadening its scope of concepts and applications.  Realized
in terms of  the AdS/CFT correspondence \cite{AdSCFT}, it asserts that gravitational dynamics in an asymptotically
Anti de Sitter (AdS) spacetime can be mapped onto a (relativistic) conformal field theory (CFT) in one less dimension. It has been a useful tool in improving our understanding 
understanding the behaviour of strongly interacting field theories (for example transport properties of QCD quark-gluon plasmas \cite{AdSquark}) since their dual description is in terms of weakly coupled gravitational dynamics in a bulk spacetime that is asymptotically AdS.  More recently this correspondence has been shown to be applicable to a class of strongly correlated  electron and atomic systems that exhibit relativistic dispersion relations (whose dynamics near 
a critical point is well described by a relativistic CFT)  and has  been used to study superconductivity, the quantum Hall effect, and a number of other condensed matter systems that can be described by CFTs \cite{AdSCmat}.   

Even more recent has been the emergence of Galilean holography,  which posits a duality between a non-relativistic 
 CFT in $d$ spatial dimensions at strong coupling and a gravitational system in $(d+3)$ dimensions  that asymptotically approaches a Schr¬odinger spacetime \cite{Galhog}. This latter spacetime
has an isometry group obtained from the well-known Galilean algebra of translations, rotations, and boosts, to include dilatations and special conformal transformations (valid when the time coordinate scales like the square of a spatial coordinate) \cite{NishSon}.  Galilean holography is expected to provide  important information about strongly coupled non-relativistic systems in the real world, such as a system of fermions 
at unitarity, which is of considerable interest in the context of trapped cold atoms at a Feshbach resonance.   Such cold atom systems may be an example of an ultra-low viscosity fluid, similar to a quark-gluon plasma \cite{coldatom}.

Finite-temperature generalizations of Galilean holography were recently proposed \cite{hotgal1,hotgal2}, in which
 a planar (toroidal) black brane solution with Schr\"odinger spacetime  asymptotics was
proposed as the holographic dual of the non-relativistic CFT at finite temperature.
To obtain this solution  a Null Melvin Twist (NMT) \cite{hotgal2,NMT} was employed on a 
planar Schwarzschild-AdS black hole (times $S^5$, thereby describing the near-horizon geometry of D3-branes in string theory) to generate the solution, whose final form was given after Kaluza-Klein (KK) reduction of the $S^5$. Thermodynamic and transport properties of this `Planar Galilean Black Hole' (PGBH) were also determined, and it was shown  no Hawking-Page transition takes place \cite{hotgal1}.  

I demonstrate here that this conclusion crucially depends upon the choice of background, and that in fact a phase transition can take place between a PGBH and a new soliton solution  that I shall call the Planar Galilean Soliton (PGS).   This latter solution can be obtained by applying an NMT to the AdS soliton \cite{AdSsol}, which is conjectured to be the lowest energy solution for spacetimes that have the same asymptotic behaviour (which includes AdS black holes with Ricci flat horizons). It is natural to regard the PGS as the thermal background since (as I will demonstate) the PGS has lower energy than the PGBH. Given its origins  it  is also reasonable to conjecture that this PGS is the lowest energy solution for spacetimes with the same asymptotic behaviour.   

AdS planar black holes are part of a more general class of AdS topological black holes \cite{AdStop}, in which
the horizon of the black hole has positive, negative or zero curvature. Being static, a standard equilibrium thermodynamics analysis can be applied to them \cite{toptemp}, and it has been shown that 
AdS black holes with Ricci flat horizons (which include the planar black holes) do not undergo phase transitions  when the zero mass black hole is taken as the thermal background \cite{nophase}.  An analogous conclusion holds for the PBGH when the zero-mass solution is taken to be the background \cite{hotgal1,Yamada}. A phase transition has been also shown to occur for Galilean black holes
with spherical topology in the transverse directions in the background of their zero-mass counterparts \cite{Yamada}.  However the results there are contingent upon an ad-hoc matching condition.  As I will demonstrate these pitfalls
can be avoided for the PGBH using the boundary counterterm approach.  

The PGBH-PGS phase transition is similar to that of a planar black hole to an AdS soliton \cite{SKD} in that
it depends on two parameters.
However unlike this case the additional
parameter for the PGBH phase transition (and  free energy),  namely the chemical potential for the particle number (realized as momentum in the light-cone direction), is given in terms of the size of its soliton background.  
 Recall that for  spherical AdS black holes in 4 or more dimensions,  the phase transition depends only on the temperature \cite{HP}. Although I shall explicitly work only in $d+3=5$ dimensions, this type of phase transition 
will take place in any number of dimensions.  

The 5D PGBH has metric, scalar $\Phi$ and gauge field $A_\mu$ given by
\begin{eqnarray}
ds^2 &=& \frac{r^2  }{K_b^{\frac{2}{3}} R^2}\left[ -(1+b^2r^2)V_b dt_b^2 -2 R b^2r^2 V_b dt_b d\phi_b   \right. 
   \nonumber\\
&& \left. + R^2 (1-b^2r^2 V_b)d\phi_b^2 + R^2 K_b d{{\mathbf{x}}^2_b} \right] 
+\frac{K_b^{\frac{1}{3}} R^2 }{r^2}\frac{dr^2}{V_b} \nonumber \\
\Phi &=& -\frac{1}{2} \ln K_b    \qquad 
A = \frac{b r^2}{K_b R^2} \left(V_b dt + d\phi_b\right) \label{e1}
\end{eqnarray}
where $V_b=  1 -  \frac{r^4_b}{r^4}$  and $K_s = 1+ \frac{b^2 r^4_b}{r^2}$. 
There is an event horizon at $r=r_b$, where $V_b(r_b)=0$.   Applying an NMT to the AdS soliton yields the PGS solution
\begin{eqnarray}
ds^2 &=& \frac{r^2  }{K_s^{\frac{2}{3}} R^2}\left[ -(1+b^2r^2 V_s) dt_s^2 -2 R b^2r^2 V_s dt_s d\phi_s   \right. 
   \nonumber\\
&& \left. + R^2 (1-b^2r^2)V_s d\phi_s^2 + R^2 K_s d{{\mathbf{x}}^2_s} \right] 
+\frac{K_s^{\frac{1}{3}} R^2 }{r^2}\frac{dr^2}{V_s} \nonumber \\ 
\Phi &=& -\frac{1}{2} \ln K_s    \qquad 
A = \frac{b r^2}{K_s R^2} \left( dt + V_s d\phi_s\right) \label{e2}
\end{eqnarray}
where now $V_s=  1 -  \frac{r^4_s}{r^4}$  and $K_s = 1- \frac{b^2 r^4_s}{r^2}$.  There is no longer an event horizon, but regularity forces  $r\geq r_s$  and $\Delta \phi_s =\frac{\pi R}{r_s}$ , where $V_s(r_s)=0$.  Furthermore $b r_s <1$ to ensure a proper signature.

Both metrics are solutions of the field equations that follow from variation of the action
\begin{equation}
I = \frac{1}{16\pi G_{5}}\left[\int d^{5}x \sqrt{-g} {\cal L}_B + \int d^{4}x \sqrt{h} {\cal L}_S\right]
\label{eqact}
\end{equation}
where
\begin{eqnarray}
{\cal L}_B &=& R - \frac{4}{3}(\nabla\phi)^2
-\frac{ e^{ -8\Phi/3}}{4} F_{\mu\nu}F^{\mu\nu} - 4 A_\mu A^\mu  -V(\Phi)  \nonumber \\
{\cal L}_S &=& 2{\cal K} - 6 + A_a A^a  - \frac{3}{2} A_a A^a \Phi \label{eqlag}
\end{eqnarray}
with  $ V(\Phi) = 4 e^{8\Phi/3}  - 16 e^{2\Phi/3}$ and ${\cal K}$ the extrinsic curvature of
the boundary whose induced metric and gauge field are respectively $h_{ab}$ and $A_a$.  Both solutions (\ref{e1},\ref{e2}) uplift to solutions of 10-dimensional Type IIB  supergravity, though there is no argument that the bulk action (\ref{eqact}) is a consistent truncation
of the 10-dimensional theory \cite{hotgal1,hotgal2}. 

The metrics, scalars, and gauge fields are asymptotically of the same form.  The metrics in solutions (\ref{e1}) and (\ref{e2}) at large $r$ are asymptotic to a metric of the form
\begin{equation}
ds^2 =  r^2 \left[ -2 du dv -  r^z du^2  + d{{\mathbf{x}}^2_b} \right] 
+\frac{R^2 dr^2}{r^2}  \label{eq3}
\end{equation}
where $z=2$, which can be obtained from either metric by writing  $R u = b(t +R\phi)$ and $2bR v = (t -R\phi)$
in the large $r$ limit. The only possible distinction between the two asymptotic solutions is in the periodicities of the $(\phi ,{\mathbf{x}}_{i})$ coordinates, which must be matched for a consistent comparison.  The asymptotic metric (\ref{eq3}) for $z=2$ exhibits the full Galilean
conformal invariance (realizing the Schr¬odinger algebra), which implies that under a scaling
transformation $r \to \lambda^{-1} r$ ,  $v$ is invariant and $u\to \lambda^2 u$ (and ${\mathbf{x}}_{i}\to \lambda {\mathbf{x}}_{i}$) .

The boundary Lagrangian (\ref{eqlag}) consists of the usual Gibbons-Hawking term plus a minimal (non-unique) set of counterterms needed to make the variational principle well-defined and the
(Euclidean) action finite.  To study the thermodynamics of the PGBH in the PGS background, it is necessary to compute the difference in the Euclidean action, $\Delta I = I_{\textrm\tiny PGBH} - I_{\textrm\tiny PGS}$,  between the two solutions.   This can be done using either the action (\ref{eqact})
or (more tediously) by retaining only the bulk and Gibbons-Hawking terms and matching the metrics on a large-$r$ boundary that is then taken to infinity after computing the difference in the actions. The result is the same in each case and yields
\begin{equation}
\Delta I =  \frac{\Delta{\mathbf{x}}^2}{16\pi G_{5}} \left(\Delta\phi_s \Delta\tau_s \frac{r^4_s}{R^3}
-  \Delta\phi_b \Delta\tau_b \frac{r^4_b}{R^3}\right) \label{eqI}
\end{equation}
where the periodicities in the ${\mathbf{x}}$ coordinates have been matched.  A consistent comparison requires that the periodicity of the Euclidean times $\Delta\tau$ and angles 
$\Delta\phi$ also match.  Regularity of the analytically continued PGBH forces $\Delta\tau_b = \frac{\pi R^2}{r_b}$ and regularity of
the soliton was already shown to determine $\Delta\phi_s$.  

Hence, writing $V=\Delta{\mathbf{x}}^2$,
\begin{equation}
\Delta I =  \frac{\pi RV}{16  G_{5}}  \left(\frac{r^3_s}{r_b} - \frac{r^3_b}{r_s}\right)
\end{equation}
a result which is the formally the same as that for the planar black hole.  However the thermodynamics is distinct for two reasons.  First, the generator of time translations for the
non-relativistic CFT is given by $\frac{\partial}{\partial u}$ in the light cone coordinates of
(\ref{eq3}), and so the normalization of the Killing generator $\xi$ of the horizon is $\frac{R}{b}\frac{\partial}{\partial t} = \frac{\partial}{\partial u} + \frac{1}{2b^2}\frac{\partial}{\partial v}$, yielding
a temperature $T$ and chemical potential $\mu$
$$
T = \frac{r_b}{\pi R b} \qquad  \mu = \frac{1}{2b^2}
$$
for the PGBH, where $\mu$ is conjugate to the momentum in the $v$ direction.  

Second, compactification in the $v$ direction, which affords a discrete light-cone quantization description of the CFT
\cite{hotgal1,DLCQ} and breaks no symmetry, implies that the chemical potential is determined in terms of
the soliton `size' $r_s$. The period of compactification $|\Delta v| = 1/m_G$ is thus a physical parameter, interpreted as the inverse of the Galilean mass in the non-relativistic CFT.  
Since $b|\Delta v| = \Delta \phi $ at fixed
$u$, this period is determined in terms of the soliton parameter $r_s$, with 
\begin{equation}
b r_s|\Delta v| = \pi R \label{eq4}
\end{equation}
and so the soliton background induces  a maximum value of $m_G < 1/\pi R$ for the Galilean mass. 

Rewriting the action in terms of $T$, $\mu$ and $m_G$ yields
\begin{equation}
F = -\frac{\pi^3 R^3 V T^4}{64 m_G G_{5} \mu^2}  \left[1 - \left(\frac{2\mu m_G  }{T}\right)^4 \right]
\label{eq5}
\end{equation}
for the free energy $F= T\Delta I$.  This  is consistent with the form of the free energy
for a scale-invariant theory at finite temperature and chemical potential, which
can be written as
\begin{equation}
F = -V T^\alpha f\left(\frac{\mu}{T}\right)
\label{eq5a}
\end{equation}
where $\alpha=1+d/z$ is determined by dimensional analysis, with $d=z=2$ for the PGBH/PGS.
This value implies that $zE = dPV$, in agreement with the scale-invariant
Ward identity \cite{hotgal2}.  Note  that $z=2$ for free nonrelativistic gases (classical and quantum with either statistics) in the grand canonical ensemble \cite{LL}.

It is clear that $F$ is not always negative, signalling a phase transition. At high temperatures the PGBH is stable, but a cold PGBH is unstable to
decay to a cold PGS. The PGBH and PGS are in equilibrium 
whenever $T=2  \mu m_G$.    This situation is similar to the Hawking-Page transition \cite{HP} (in which
high temperature black holes are stable while low temperature black holes decay to  global AdS spacetime), and opposite to that of AdS black holes with Ricci-flat horizons \cite{SKD} (in which small hot black holes are unstable to decay to small hot AdS solitons).  The distinction can be traced back to the constraint 
(\ref{eq4}) that relates the compactification scale (and in turn the chemical potential) to the size of the soliton, a condition not present for AdS black holes with Ricci-flat horizons.

Interpreting the action (\ref{eqI}) as the saddle-point approximation to the grand canonical
partition function, $Z(T,\mu) \approx \exp(-I)$, 
it is straightforward to compute the entropy $S$, particle number $N=P/(2\pi m_G)$, and energy $E$  via the relations
\begin{eqnarray}
S = -\frac{\partial F}{\partial T}   &\qquad&  P = \frac{\partial F}{\partial \mu}  \\
 &E = F+TS-\mu P  \label{eq6}&
\end{eqnarray}
yielding
\begin{eqnarray}
S &=&  \frac{\pi^3 R^3 V T^3}{16 m_G G_{5} \mu^2}  \\
N &=& \frac{\pi^2 R^3 V T^4}{64 m^2_G G_{5} \mu^3} \left[1 + \left(\frac{2 \mu m_G}{T}\right)^4 \right] \\
E &=& \frac{\pi^3 R^3 V T^4}{64 m_G G_{5} \mu^2}  \left[1 - \left(\frac{2 \mu m_G }{T}\right)^4 \right]
\end{eqnarray}

The entropy S can be shown to be equal to $1/(4 G_5)$ times the area of the event horizon
of the PGBH, providing a consistency check on the results.  As promised the energy of the
PGBH relative to the soliton is always positive in the black hole phase.  It is straightforward to show that the first law of thermodynamics $dE = T dS -\mu dP$ holds for this solution. The specific heat 
\begin{equation}
C = \left. \frac{\partial E}{\partial T}\right|_\mu = \frac{\pi R^3 V T^3}{16 m_G  G_{5} \mu^2} 
\end{equation}
is always positive, increasing linearly with the black hole area. Since the pressure ${\cal P} = -Z/V $ in the grand canonical ensemble, one obtains
\begin{equation}
PV = E
\end{equation}
which is a relation that holds for a non-relativistic system with Galilean conformal invariance
with two transverse spatial dimensions as noted above. 

At any fixed temperature $T$ the particle number has a minimum value  and so
\begin{equation}
N \geq N_{\textrm{\tiny min}} = 3^{1/4} \frac{\pi^2 R^3 VT  m_G }{6 G_{5}  } 
\end{equation}
which occurs at at $\mu = 3^{1/4} T/(2m_G)$ below the phase transition.
The computation of the viscosity from the Kubo formula yields the expected relation
\begin{equation}
\eta = \frac{S}{4\pi V}
\end{equation}
as obtained by a consideration of fluctuations of the black hole metric (\ref{e1}) in the spatial directions of the nonrelativistic field theory \cite{hotgal1}.

The PGBH should correspond to the high-temperature phase of  the Galilean CFT, whereas
the PGS should correspond to its low temperature phase.  Although there has been an expectation that non-relativistic CFTs in finite volume should exhibit a phase transition similar to the Hawking-Page transition, there is no need to compactify the transverse $\mathbf{x}$ directions here.  The phase transition instead is a consquence of
the lower energy of the soliton, which also sets the compactification scale. 

A consideration of more general Galilean black holes with Ricci-flat horizons indicates that the above results will hold in this case as well.  The free energy will be of the form
\begin{equation}
F = -\frac{\pi^{d+1} R^{d+1} V_d T^{\frac{d+2}{2}}}{2^{\frac{d+10}{2}}m_G  G_{d+3}}\left(\frac{T}{\mu} \right)^{\frac{d+2}{2}}  \left[1 - \left(\frac{2 \mu m_G }{T}\right)^{d+2} \right]
\label{eq5}
\end{equation}
in $(d+3)$ dimensions, which follows from straightforward dimensional analysis.  Hence all such black holes will be unstable to decay into solitons at sufficiently low temperature.

Since the Euclidean temporal $S^1$ of the PGBH is contractible, it can support only anti-periodic boundary conditions on spinor fields, whereas this factor is non-contractible for the PGS and so it can support both periodic and anti-periodic spinorial boundary conditions.  Hence this phase transition can take place only for the anti-periodic partition function Tr[$\exp(-H/T)$].  This argument is analogous to that for spherical AdS black holes \cite{AdSCFT} and for 
AdS Black holes with Ricci flat horizons \cite{SKD}.

The PGBH/PGS phase transition does not demonstrate that the PGS spacetime is the preferred ground state, with  minimal energy in its topological class.  Rather it only demonstrates that some phase transition
at low temperature must take place, provided the appropriate boundary conditions are satisfied.  Whether this spacetime is of the PGS type or something else remains
an interesting open question.

\section*{Acknowledgements}
I am grateful to D. Marolf and M. Rangamani for helpful discussions, and to the Kavli Institute for Theoretical Physics where this work was carried out.  This work was supported by the Natural Sciences \& Engineering Research Council of Canada and by the Fulbright Foundation.

\end{document}